# Magnetic-Field Tuning of Light-Induced Superconductivity in Striped $La_{2-x}Ba_xCuO_4$

D. Nicoletti[1,*], D. Fu[1], O. Mehio[1], S. Moore[1], A. S. Disa[1], G. D. Gu[2], and A. Cavalleri[1,3]

[1] *Max Planck Institute for the Structure and Dynamics of Matter, 22761 Hamburg, Germany*
[2] *Condensed Matter Physics and Materials Science Department,*
*Brookhaven National Laboratory, Upton, New York 11973, USA*
[3] *Department of Physics, Clarendon Laboratory, University of Oxford, Oxford OX1 3PU, United Kingdom*
*\* e-mail: daniele.nicoletti@mpsd.mpg.de*

## Abstract

*Optical excitation of stripe-ordered $La_{2-x}Ba_xCuO_4$ has been shown to transiently enhance superconducting tunneling between the $CuO_2$ planes. This effect was revealed by a blue-shift, or by the appearance of a Josephson Plasma Resonance in the terahertz-frequency optical properties. Here, we show that this photo-induced state can be strengthened by the application of high external magnetic fields oriented along the c-axis. For a 7-Tesla field, we observe up to a ten-fold enhancement in the transient interlayer phase correlation length, accompanied by a two-fold increase in the relaxation time of the photo-induced state. These observations are highly surprising, since static magnetic fields suppress interlayer Josephson tunneling and stabilize stripe order at equilibrium.*

*We interpret our data as an indication that optically-enhanced interlayer coupling in $La_{2-x}Ba_xCuO_4$ does not originate from a simple optical melting of stripes, as previously hypothesized. Rather, we speculate that the photo-induced state may emerge from activated tunneling between optically-excited stripes in adjacent planes.*

Charge and spin ordered phases are found throughout wide regions of the phase diagrams of high-$T_C$ cuprates [1,2,3,4,5,6,7]. These orders tend to compete with the superconducting state, reducing its coherence. A well-studied case is that of single-layer compounds, in which the doped holes arrange themselves in one-dimensional charge stripes separated by regions of oppositely phased antiferromagnetic order in the $CuO_2$ planes. In $La_{2-x}Ba_xCuO_4$ (LBCO), for example, stripes completely suppress superconductivity at $x$ = 1/8 doping [8,9], and coexist with it at lower and higher doping values (see phase diagram in Fig. 1(a)).

Recent experimental evidence [10,11] suggests that the individual striped planes may be made up of a spatially modulated superfluid, a so-called "pair-density-wave" (PDW), in which the interlayer Josephson tunneling is frustrated by symmetry (see 90° stacking in inset of Fig. 1(a)) [12,13,14].

Pressure and magnetic fields have been used to tune the striped state in LBCO and in related compounds. In particular, hydrostatic pressures of few GPa were shown to increase $T_C$ [15] and to partially suppress charge order [16]. On the other hand, magnetic fields $H \lesssim$ 10 T were shown to amplify the effect of dynamical layer decoupling, leading to a reduction of interlayer tunneling [17,18] and the stabilization of charge- and spin-order [19,20].

More recently, optical excitation with femtosecond laser pulses has emerged as a means to drive the interplay between stripes and superconductivity [21], enhancing one or the other transiently. Excitation of either the in-plane Cu-O stretching mode in non-superconducting $La_{1.675}Eu_{0.2}Sr_{0.125}CuO_4$ (LESCO$_{1/8}$) [22,23], or of $La_{2-x}Ba_xCuO_4$ with high-energy (1.5 eV) optical pulses [24,25], were both shown to enhance interlayer tunneling [26].

In these photo-induced superconductivity experiments, the enhancement was achieved at or near 1/8 doping levels, and it was generally interpreted as a consequence of the removal of frustration by ultrafast melting of stripes [27,28].

Here, we study this effect by tuning the relative strength of equilibrium interlayer tunneling and stripes before photo-excitation. This is investigated for different doping levels, as a function of temperature, and in presence of external magnetic fields up to 7 T. The experiments yield highly surprising results and identify a correlation between the strength of the equilibrium stripe phase (before it is optically melted) and the photo-induced superconducting state.

The $La_{2-x}Ba_xCuO_4$ single crystals used in our experiments were grown at two nominal Ba concentrations $x$ = 9.5% and 11.5% [9]. Both of these compounds are superconducting, with transitions at $T_C \simeq$ 32 and 13 K, respectively. At 9.5% doping, superconductivity, charge- and spin-order all appear at the same temperature $T_{CO} \simeq T_{SO} \simeq T_C \simeq$ 32 K, while these transitions become decoupled at higher hole concentrations ($T_{CO} \simeq$ 53 K and $T_{SO} \simeq$ 40 K for $x$ = 11.5%, see phase diagram in Fig. 1(a)).

All measurements were carried out in a superconducting magnet with optical access, with magnetic fields up to 7 T applied along the $c$ direction, and for temperatures down to ~5 K (see Fig. 1(b) for experimental geometry).

The equilibrium optical properties [29] at both doping levels were determined using single-cycle THz pulses polarized along the $c$ axis (see Fig. 1(b)), whose electric field profile was measured after reflection from the sample surface at different temperatures, both below and above $T_C$ [30]. The $c$-axis equilibrium reflectivities extracted with this procedure are shown in the left-hand panel of Fig. 1(c) for $H$ = 0 and $T$ = 5 K. Both spectra are characterized by a Josephson Plasma Resonance (JPR), which appears at ~0.5 THz for $x$ = 9.5% and at ~0.2 THz for $x$ = 11.5%.

In the time resolved experiments, the LBCO crystals were photo-excited with ~100 fs, 800 nm wavelength laser pulses, also polarized along the *c* axis (see Fig. 1(b)), at a fluence of ~2 mJ/cm². The pump photon energy is indicated by arrows in the spectra of Fig. 1(c). Changes in the real and imaginary *c*-axis optical properties were retrieved for different time delays after photo-excitation with a temporal resolution of ~350 fs [39], having accounted for the pump-probe penetration depth mismatch [30].

The optical response of the 11.5%-doped compound ($T < T_C$), measured in absence of magnetic field, is displayed in Fig. 2(a-d), both at equilibrium (gray) and at two different pump-probe time delays after photo-excitation (blue circles). As already reported in Refs. [24,25], this material was found to be the most photo-susceptible within the LBCO family, displaying optically-enhanced (induced) superconductivity all the way up to $T_{SO} \simeq 40$ K.

The τ = 1.4 ps data show a response compatible with a strong optically-induced increase in interlayer tunneling strength, evidenced by a blue-shift of the reflectivity edge from ~0.2 THz to ~0.6 THz (see Fig. 2(a)), and, correspondingly, by an enhancement in the low-frequency imaginary conductivity $\sigma_2(\omega)$ (Fig. 2(c)) [30]. The data shown in Fig. 2(a) and Fig. 2(c) could be fitted using a model that describes the optical response of a Josephson plasma, for which the dielectric function is expressed as $\tilde{\varepsilon}(\omega) = \varepsilon_\infty\left(1 - \omega_J^2/\omega^2\right)$. Fits to the transient spectra are displayed as blue lines [30].

At longer time delay (τ = 2 ps), the response evolved into that of a conductor with finite momentum relaxation rate, as evidenced by the broadening of the reflectivity edge (Fig. 2(b)) and by a downturn of $\sigma_2(\omega)$ at low frequency (Fig. 2(d)) [30]. Notably, the response functions at this longer time delay were fitted using a conventional Drude model for metals (blue lines in Fig. 2(b) and Fig. 2(d)), for which the complex dielectric function is expressed as $\tilde{\varepsilon}(\omega) = \varepsilon_\infty[1 - \omega_P^2/(\omega^2 + i\Gamma\omega)]$. Here, $\omega_P$ and $\Gamma$ are the carrier plasma frequency and momentum relaxation rate, respectively [30].

The same experiment was repeated in a 7 T magnetic field, oriented along the c axis. At equilibrium, the application of the magnetic field induces vortices within the CuO$_2$ layers, stabilizes charge and spin order [20], and suppresses interlayer Josephson coupling [17] without appreciably affecting superconductivity in the planes (as $H_{c2} \gg 7$ T). The magnetic-field-induced suppression of interlayer superfluid stiffness is evidenced by the red-shifted equilibrium JPR, which for zero field is found at ~0.2 THz (gray line in Fig. 2(a)), while in a 7 T magnetic field it is quenched to below the probed frequency range (< 0.15 THz, gray line in Fig. 2(e)).

Similar to the data of Fig. 2(a), also in this case a clear edge appeared in the transient reflectivity measured 1.4 ps after photo-excitation (Fig. 2(e), red circles). However, here the transient plasma resonance remained sharp for longer time delays (Fig.2(f)), revealing superconducting-like optical properties even at τ = 2 ps. This is also evidenced by the diverging $\sigma_2(\omega \to 0)$ in Fig. 2(h) [30]. Remarkably, unlike the zero-field data in Fig. 2(a-d), the transient optical properties measured at 7 T could be fitted with the Josephson plasma formula at both time delays shown in Fig. 2(e-h).

The coherent character of this newly discovered optically-enhanced superconducting state in presence of high magnetic fields is even more evident in Fig. 3. Here, we show, for different data sets, the dynamical evolution of the interlayer phase correlation length, extracted from the Drude fits of Fig. 2 and defined as $\xi_c = 2\omega_P L/\Gamma$ (here $L$ is the CuO$_2$ layer separation) [23]. Note that at the earliest time delays, for which the optical properties could also be fitted with a JPR model, $\xi_c$ is at least ~20 unit cells. The upper bound is here determined by the frequency resolution of our measurement and it corresponds to carrier mobilities larger than ~10$^4$ cm$^2$/(V·s). Importantly, these are extremely high values, which would be unprecedented for out-of-plane transport in a highly resistive normal state oxide, and are rather strongly suggestive of a transient superconducting state (see Section S4 in Supplemental Material [30] for an extended discussion on this topic).

At longer time delays, for which the response is well described by the Drude model, $\xi_c$ becomes finite and exhibits a collapse, within a few picoseconds, down to values close to 1-2 unit cells. This collapse became slower and not faster when the temperature was raised to 30 K (Fig. 3(a)), and it was observed to be even slower in presence of a 7 T magnetic field (Fig. 3(b)). Notably, the application of magnetic field resulted, for certain time delays ($\tau \simeq$ 1-2 ps) in a ~10-fold increase in $\xi_c$.

A relevant piece of information is added here by the measurements performed at $T_C < T < T_{SO}$ (Fig. 3(a)) [30]. In this case, transient superconductivity emerged from the equilibrium normal state, displaying a larger $\xi_c$ than that found at lower temperatures. This observation suggests that, for $H$ = 0, the transient superconductor induced from the normal state is "more coherent" than that induced by exciting the superconducting state at $T < T_C$.

Importantly, all data reported in Fig. 3 display relaxation dynamics to a state with only partially reduced coherence. At $\tau \gtrsim$ 2 ps, the data could still be fitted with momentum relaxation rates of $\Gamma \simeq$ 1 THz (corresponding to $\xi_c \sim$ 1-2 unit cells, see also Ref. [24]), values that are anomalously low for conventional incoherent charge transport, and are instead compatible with a strongly fluctuating superconducting state [40].

The full dynamical response, determined for a wide variety of initial conditions (different doping values, temperatures, external magnetic fields) is shown in Fig. 4. Here, we plot the frequency- and time-delay-dependent energy loss function, $\Im[-1/\tilde{\varepsilon}(\omega,\tau)]$, as extracted from fits to the transient optical response functions. This quantity exhibits a sharp peak at the JPR frequency, which acquires a finite width Γ as soon as momentum relaxation processes set in.

Figure 4(a) displays the response of the 9.5%-doped compound [30] below $T_C$ at zero field, that was already reported in Ref. [24]. This material, for which the equilibrium superconducting phase is robust (see equilibrium JPR at ~0.5 THz) and stripe order is weak [9], only shows marginal superconductivity enhancement for $\tau \lesssim 1$ ps, and then, at later delays, it evolves abruptly into an incoherent state, characterized by an overdamped loss function peak.

When a 7 T magnetic field is applied to the 9.5% compound (Fig. 4(b)), the equilibrium interlayer superfluid stiffness is reduced (the JPR at $\tau \leq 0$ ps is now at ~0.2 THz) and, concomitantly, stripe order is enhanced by ~30% [20]. Here, the optically-induced effect was different from that measured in the same material at $H = 0$ and resembled that found at 11.5% doping. Within ~1.5 ps after photo-excitation, a notable blue-shift of the equilibrium JPR developed, reaching values close to 0.5 THz [30]. This observation suggests that interlayer Josephson tunneling, which was almost completely quenched by the magnetic field, can be transiently revived also at 9.5% doping by photo-excitation. However, this revival occurs only over a short time interval, and for $\tau \gtrsim 1.5$ ps the system evolves abruptly into an incoherent state.

The middle and lower panels of Fig. 4 (panels c-f) show the $\Im\mathfrak{m}[-1/\tilde{\varepsilon}(\omega,\tau)]$ function of the 11.5% compound, for which selected optical spectra are displayed in Fig. 2 and the extracted phase correlation lengths are reported in Fig. 3 [30]. Note that this compound is characterized by weaker equilibrium interlayer superfluid stiffness and by a far stronger stripe order than that found at 9.5% doping (by a factor of ~5) [9].

The dynamical evolution of the loss function in this compound displays, for all temperatures and applied magnetic field values, a more "coherent" character than that found in La$_{1.905}$Ba$_{0.095}$CuO$_4$ (Fig. 4(a-b)). In particular, the data taken at T = 5 K and $H$ = 7 T (Fig. 4(f)) are those showing the sharpest resonance, with the longest lifetime.

All results above clearly indicate that optically-enhanced (induced) superconductivity in La$_{2-x}$Ba$_x$CuO$_4$ correlates with the strength of the equilibrium stripe order. Indeed, the stronger the stripes, the longer the photo-induced coherence, as shown by systematically changing the doping level, sample temperature and external magnetic field. An exception to this trend is found in the temperature dependent response of LBCO 11.5% (Fig. 3(a) and Fig. 4(c-d)). Here, the transient lifetime is clearly enhanced in the 30 K data, although the equilibrium charge order is reduced by ~15% with respect to lower temperatures [9].

The observations above can be interpreted by considering the role of the equilibrium, pre-existing *c*-axis superfluid stiffness in determining the properties of the transient state in the LBCO compounds investigated in our study. Our data suggest that a more robust superconductor at equilibrium adversely affects the strength and lifetime of the photo-induced interlayer coupling (see Fig. 4(a)), while a weak or completely absent pre-existing condensate (Fig. 4(d-f)) promotes the transient superconducting state.

We speculate that the equilibrium and optically-induced interlayer tunneling might indeed originate from two separate entities. The former is a spatially homogeneous in-plane condensate, which contributes to the interlayer superfluid stiffness at equilibrium, and whose density is higher for dopings away from 1/8 (as in LBCO 9.5%). The latter, within a PDW picture [13], may be understood as originating from superconducting stripes, whose Josephson tunneling is activated via photo-excitation and whose strength, as for the stripes, is maximum close to 1/8 doping. The equilibrium superconductor and the photo-excited PDW may well compete rather than cooperate with each other, similar to how static charge- and spin-order compete with 3D superconductivity at equilibrium.

In the aggregate, the results reported above set new constraints on the overall origin of light-induced superconductivity in single-layer cuprates, which appears to result from the photo-excited striped phase and perhaps is not easily explained as simple optical melting of stripes.

**Acknowledgments**

The research leading to these results received funding from the European Research Council under the European Union's Seventh Framework Programme (FP7/2007-2013)/ERC Grant Agreement No. 319286 (QMAC). We acknowledge support from the Deutsche Forschungsgemeinschaft via the excellence cluster 'The Hamburg Centre for Ultrafast Imaging—Structure, Dynamics and Control of Matter at the Atomic Scale' and the priority program SFB925. Work at Brookhaven is supported by the Office of Basic Energy Sciences, Division of Materials Sciences and Engineering, U.S. Department of Energy under Contract No. DE-SC0012704.

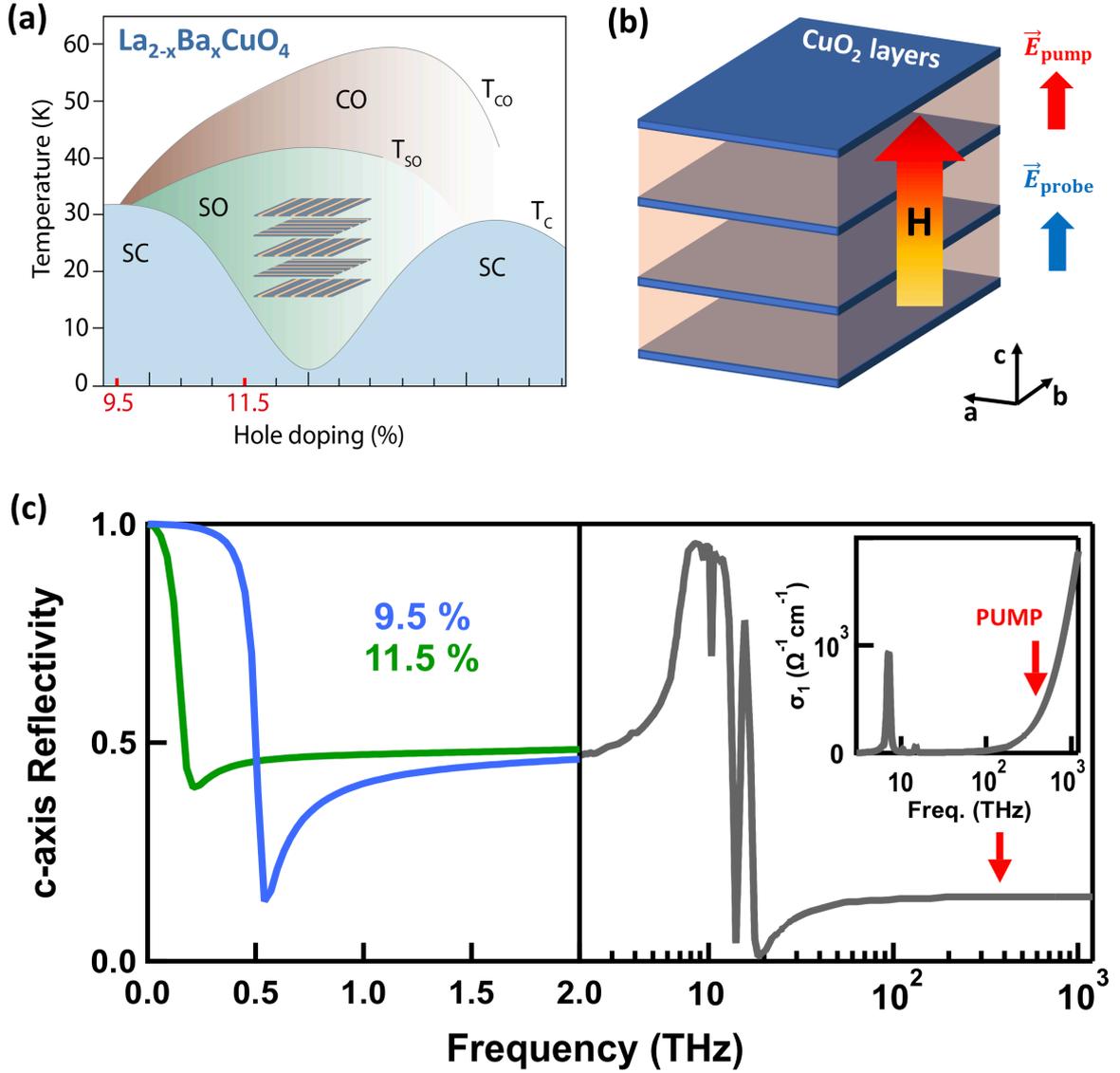

**Figure 1.** (a) Temperature-doping phase diagram of La$_{2-x}$Ba$_x$CuO$_4$ (LBCO), as determined in Ref. 9. $T_C$, $T_{SO}$, and $T_{CO}$ indicate the superconducting, spin-order, and charge-order transition temperatures, respectively. The inset shows the 90° periodic stacking of CuO$_2$ planes in the stripe phase. (b) Cartoon depicting the experimental geometry. We used optical pump and THz probe pulses polarized both along the $c$ axis. The experiment was performed in presence of magnetic fields up to 7 T, oriented along the $c$ direction. (c) Equilibrium $c$-axis optical properties of LBCO. Left panel: THz reflectivity of both samples at $T$ = 5 K (no magnetic field), resulting from fits to the experimental data [30]. Right panel and inset: broadband $c$-axis reflectivity and optical conductivity of LBCO ($x$ = 9.5%) from Ref. [29]. Red arrows indicate the pump photon energy.

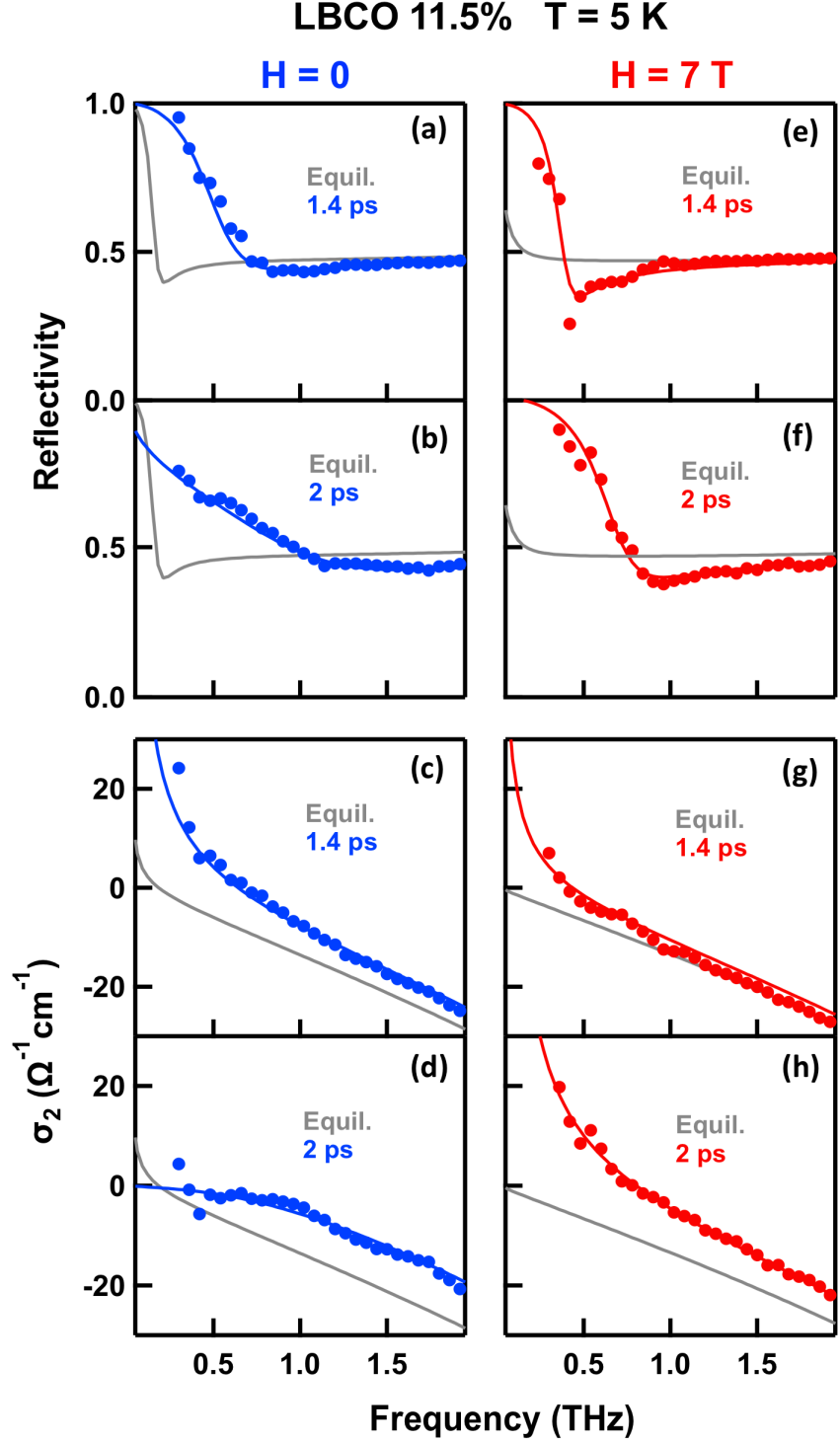

**Figure 2.** (a,b) *c*-axis THz reflectivity of La$_{1.885}$Ba$_{0.115}$CuO$_4$ measured at *T* = 5 K (no magnetic field), at equilibrium (gray lines) and at two pump-probe time delays ($\tau$ = 1.4 ps and $\tau$ = 2 ps) after excitation (blue circles). Fits to the transient spectra are shown as blue lines. (e,f) Same quantities as in (a,b) measured in presence of a 7 T magnetic field (transient data are shown here in red). (c,d) Zero-field imaginary conductivity spectra corresponding to the reflectivities shown in (a,b). (g,h) In-field imaginary conductivities corresponding to the data of (e,f).

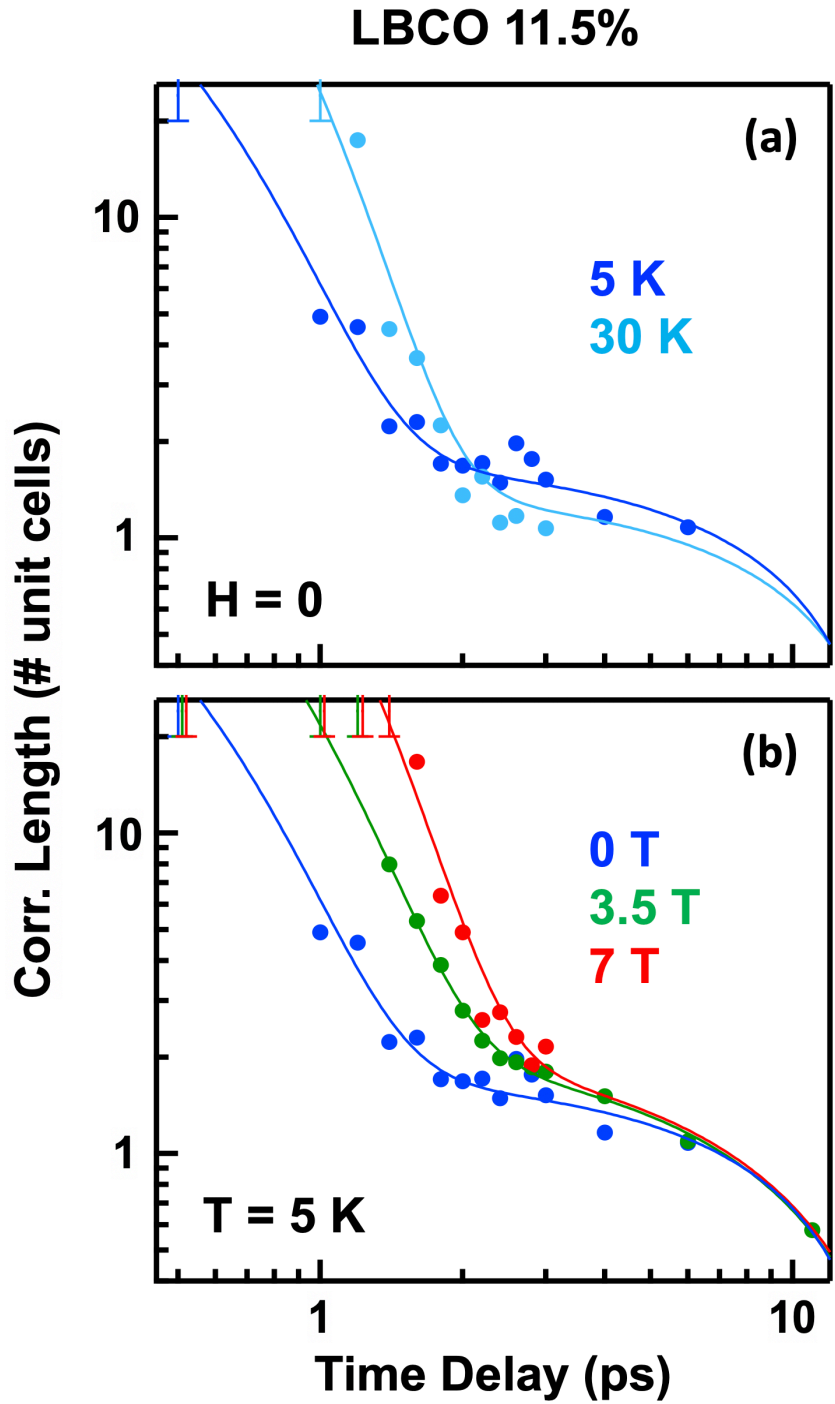

**Figure 3.** Dynamical evolution of the interlayer phase correlation length measured in La$_{1.885}$Ba$_{0.115}$CuO$_4$ for different temperatures and applied magnetic fields. The values were extracted from the fits of Fig. 2 [30].

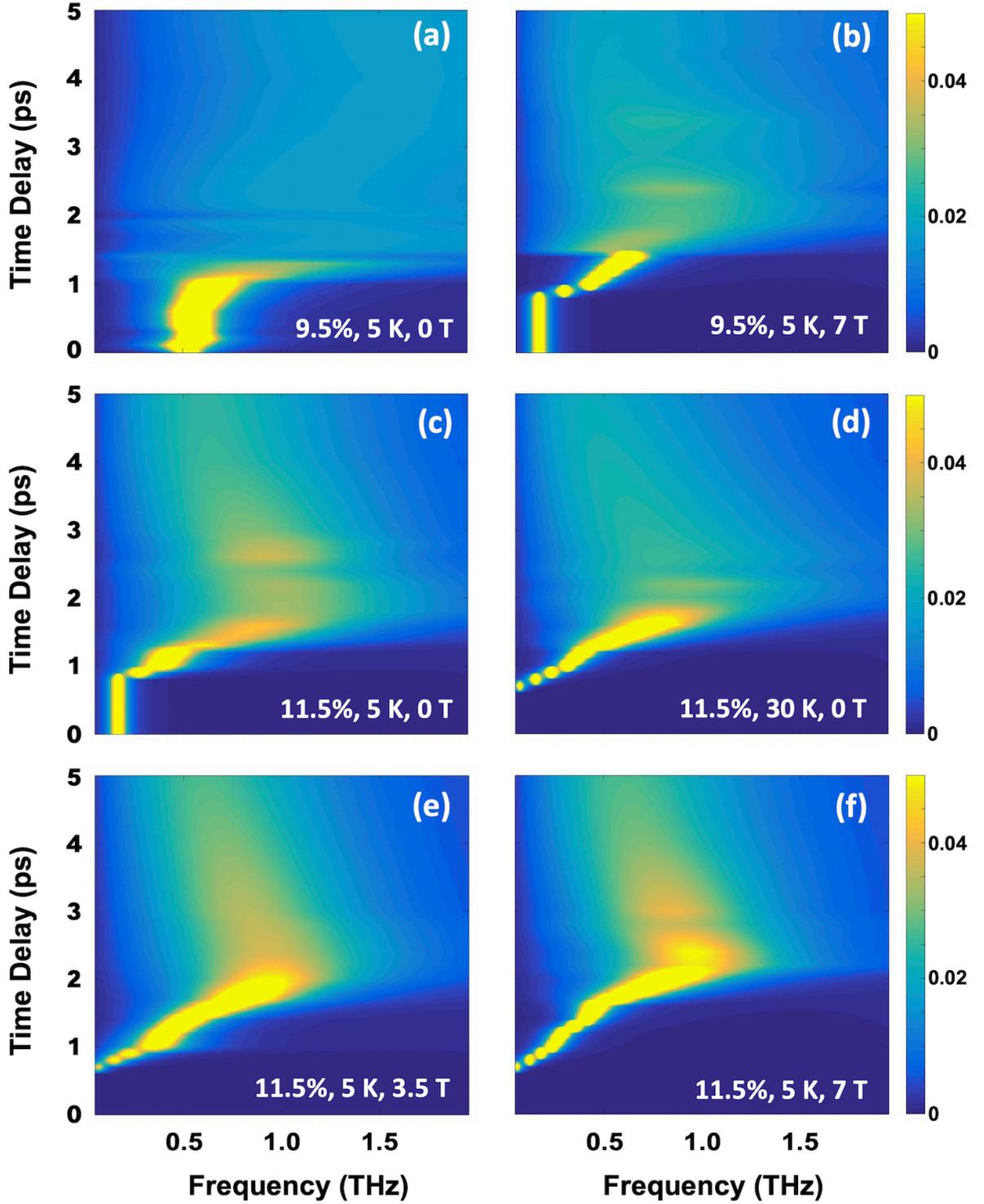

**Figure 4.** Dynamical evolution of the energy loss function of photo-stimulated LBCO under various starting conditions (different doping levels, sample temperatures, and applied magnetic field values), as extracted from fits to the data. Panels (a,b) refer to La$_{1.905}$Ba$_{0.095}$CuO$_4$ at $T$ = 5 K in absence (a) and in presence (b) of a 7 T magnetic field. Panels (c-f) show instead the response of La$_{1.885}$Ba$_{0.115}$CuO$_4$, measured at $T$ = 5 K and $H$ = 0 (c), $T$ = 30 K and $H$ = 0 (d), $T$ = 5 K and $H$ = 3.5 T (e), $T$ = 5 K and $H$ = 7 T (f) [30].

# Magnetic-Field Tuning of Light-Induced Superconductivity in Striped La$_{2-x}$Ba$_x$CuO$_4$

D. Nicoletti[1,*], D. Fu[1], O. Mehio[1], S. Moore[1], A. S. Disa[1], G. D. Gu[2], and A. Cavalleri[1,3]

[1] Max Planck Institute for the Structure and Dynamics of Matter, 22761 Hamburg, Germany
[2] Condensed Matter Physics and Materials Science Department,
Brookhaven National Laboratory, Upton, New York 11973, USA
[3] Department of Physics, Clarendon Laboratory, University of Oxford, Oxford OX1 3PU, United Kingdom
* e-mail: daniele.nicoletti@mpsd.mpg.de

# Supplemental Material

## S1. Equilibrium optical response

Both La$_{2-x}$Ba$_x$CuO$_4$ (LBCO) crystals ($x$ = 9.5% and $x$ = 11.5%) were cut and polished to give an *ac* surface with large enough area (~10 mm$^2$) to perform long-wavelength THz spectroscopy. The equilibrium optical properties at both doping levels were determined using single-cycle THz pulses generated by illuminating a photoconductive antenna with near-infrared laser pulses from a Ti:Sa amplifier. These probe pulses were focused onto the sample surface, with a ~7° incidence angle and polarization aligned perpendicular to the CuO$_2$ planes (along the *c* direction). The reflected electric field, $E_R(t)$, was measured by electro-optic sampling in ZnTe at different temperatures, both below and above $T_C$, and for various applied magnetic fields. This was then Fourier transformed to obtain the complex-valued, frequency dependent $\tilde{E}_R(\omega)$. Note that, with respect to the pump-probe measurements (see Section S2), the higher signal-to-noise ratio available for the equilibrium characterization allowed us to collect reliable data down to lower frequencies (~0.15 THz), thus covering the Josephson Plasma Resonance of the weakly coupled LBCO 11.5% compound.

The equilibrium reflectivity in the superconducting state, $R(\omega, T < T_C)$, was determined as $R(\omega, T < T_C) = \frac{|\tilde{E}_R(\omega, T<T_C)|^2}{|\tilde{E}_R(\omega, T \gtrsim T_C)|^2} R(\omega, T \gtrsim T_C)$. Here, $R(\omega, T \gtrsim T_C)$ is the normal-state reflectivity measured with Fourier-transform infrared spectroscopy on the same batch of samples [1], which is completely flat and featureless in the THz range. Two examples of equilibrium reflectivity spectra retrieved with this procedure are shown in Fig. S1 (blue and green circles for $x$ = 9.5% and $x$ = 11.5%, respectively).

These reflectivities were then fitted with the model that describes the optical response of a Josephson plasma (blue and green lines in Fig. S1, see Section S4 for fit model) and merged at $\omega \simeq 2.5$ THz with the broadband spectra from Ref. [1] (Fig. 1(c), right panel, in main text). This allowed us to perform Kramers-Kronig transformations, thus retrieving full sets of equilibrium optical response functions (*i.e.*, complex optical conductivity $\tilde{\sigma}_0(\omega)$, complex dielectric function $\tilde{\varepsilon}_0(\omega)$, complex refractive index $\tilde{n}_0(\omega)$) for all temperatures and magnetic field values investigated in our pump-probe experiment. An example of broadband equilibrium optical conductivity is reported in the inset of Fig. 1(c) in the main text.

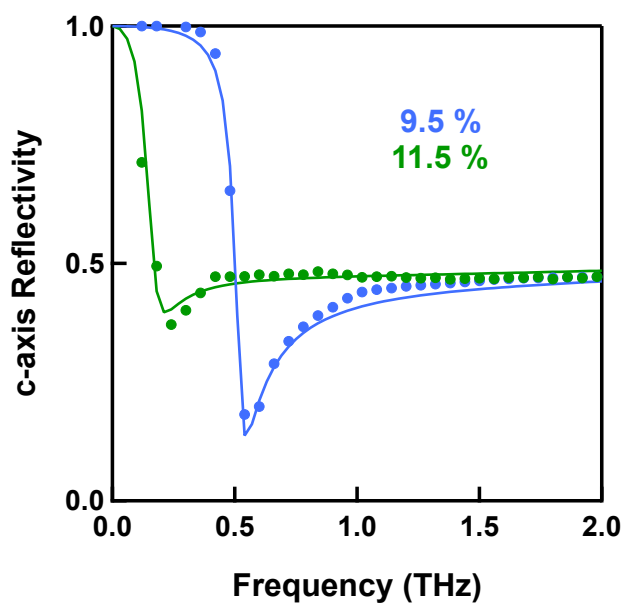

**Figure S1.** Equilibrium *c*-axis reflectivity of LBCO measured with THz time-domain spectroscopy, as described in the text. Blue and green circles are experimental data taken at *T* = 5 K (no magnetic field) on the *x* = 9.5% and *x* = 11.5% sample, respectively. Fits with a model describing the optical response of a Josephson plasma are displayed as solid lines.

## S2. Evaluation of the transient optical properties

The same THz time-domain spectroscopy geometry used for the equilibrium characterization (see Section S1) was also employed to measure the transient response of the sample after photo-excitation. Optical pulses of ~100 fs duration and 800 nm wavelength from the same Ti:Sa amplifier were shone at normal incidence onto the sample surface, with polarization along the *c* axis and a fluence of ~2 mJ/cm². Delayed THz probe pulses were used to measure the pump-induced reflectivity changes for frequencies between ~0.3 and 2.5 THz.

The pump-induced change in the THz electric field $\Delta E_R(t,\tau) = E_R^{pumped}(t,\tau) - E_R(t)$ was acquired at each time delay $\tau$ by filtering the electro-optic sampling signal with a lock-in amplifier, triggered by modulation of the optical pump with a mechanical chopper. This measurement yielded "pump on" minus "pump off" reflected electric field.

The stationary field, $E_R(t)$, was determined for each measurement by chopping the probe beam while keeping the pump *on* at negative time delay, *i.e.* as $E_R(t, \tau \ll 0)$. Here, the pump was set to hit the sample with the same power as that used for $\Delta E_R(t,\tau)$, thus accounting for possible average heating effects.

The differential electric field $\Delta E_R(t,\tau)$ and the stationary reflected electric field $E_R(t)$ were then independently Fourier transformed to obtain the complex-valued, frequency dependent $\Delta \tilde{E}_R(\omega,\tau)$ and $\tilde{E}_R(\omega)$.

Importantly, the same measurement was repeated by (i) *directly* recording $\tilde{E}_R^{pumped}(\omega,\tau)$ and $\tilde{E}_R(\omega)$ without chopping the pump and then calculating $\Delta \tilde{E}_R(\omega,\tau) = \tilde{E}_R^{pumped}(\omega,\tau) - \tilde{E}_R(\omega)$, or by (ii) acquiring $\Delta \tilde{E}_R(\omega,\tau)$ and $\tilde{E}_R(\omega)$ simultaneously at each time delay $\tau$ by filtering the electro-optic sampling signals with two lock-in amplifiers. Method (i) does not require calibration of the absolute phase of the lock-in amplifier and eliminates phase errors in estimating the optical properties, while method (ii) avoids the introduction of possible artifacts due to long term drifts and is particularly useful when the measured electric field contains fast-varying frequencies. All these methods yielded identical results.

The complex reflection coefficient of the photo-excited material, $\tilde{r}(\omega,\tau)$, was determined using the relation

$$\frac{\Delta \tilde{E}_R(\omega,\tau)}{\tilde{E}_R(\omega)} = \frac{\tilde{r}(\omega,\tau) - \tilde{r}_0(\omega)}{\tilde{r}_0(\omega)}$$

To calculate these ratios, the stationary reflection coefficient $\tilde{r}_0(\omega)$ was extracted at all temperatures from the equilibrium optical properties, determined independently at the same temperature and magnetic field value (see Section S1).

These "raw" light-induced reflectivity changes were only ~0.5-1% and needed to be reprocessed to take into account the mismatch between the penetration depth, $d(\omega) = \frac{c}{2\omega \cdot \text{Im}[\tilde{n}_0(\omega)]}$, of the THz probe [ $d(\omega \simeq 0.3 - 2.5 \text{ THz}) \simeq 50 - 200\ \mu m$ ] and that of the optical pump [ $d(\omega \simeq 375 \text{ THz}) \simeq 0.4\ \mu m$ ]. A schematic representation of the pump-probe penetration depth mismatch is displayed in Fig. S2.

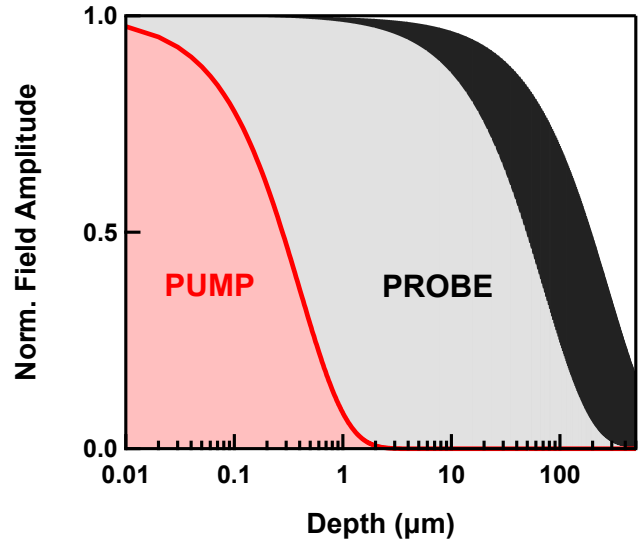

**Figure S2.** Schematics of pump-probe penetration depth mismatch. The exponential decay of both pump (red) and probe (black) field profiles is displayed as a function of depth into the material. The black region includes all field decay profiles within the bandwidth of the probe pulse.

This mismatch can be taken into account by modeling the response of the system as that of a homogeneously photo-excited layer of thickness $d \simeq 0.4$ $\mu$m with the unperturbed bulk beneath it.

The complex reflection coefficient of such multilayer system is expressed as [2]:

$$\tilde{r}(\omega,\tau) = \frac{\tilde{r}_A(\omega,\tau) + \tilde{r}_B(\omega,\tau)e^{2i\delta(\omega,\tau)}}{1 + \tilde{r}_A(\omega,\tau)\tilde{r}_B(\omega,\tau)e^{2i\delta(\omega,\tau)}}$$

where $\tilde{r}_A(\omega,\tau)$ and $\tilde{r}_B(\omega,\tau)$ are the reflection coefficients at the interfaces vacuum/photoexcited layer and photoexcited layer/unperturbed bulk, respectively, while $\delta = 2\pi d \tilde{n}(\omega,\tau)/\lambda_0$ (here $\tilde{n}(\omega,\tau)$ is the complex refractive index of the photo-excited layer and $\lambda_0$ is the probe wavelength).

The above equation can be solved numerically, thus retrieving $\tilde{n}(\omega,\tau)$ from the experimentally determined $\tilde{r}(\omega,\tau)$, and from this the complex conductivity for a volume that is homogeneously transformed,

$$\tilde{\sigma}(\omega,\tau) = \frac{\omega}{4\pi i}[\tilde{n}(\omega,\tau)^2 - \varepsilon_\infty].$$

(here $\varepsilon_\infty = 4.5$, a standard value for high-$T_C$ cuprates [3]).

The consistency of this multilayer model was also checked against an alternative method which treats the excited surface as a stack of thin layers with a homogeneous refractive index and describes the excitation profile by an exponential decay [4,5]. Both methods yielded very similar results.

Note that, in order to minimize the effects of pump-probe time resolution due to a finite duration of the probe pulse, we operated the delay stages in the setup as explained in Ref. [6]. Therefore, our temporal resolution is limited only by the duration of the pump pulse and by the inverse bandwidth of the probe pulse. For all measurements presented here the time resolution is of the order of 350 fs, the signal develops in ~1.5 – 2 ps, and the relaxation occurs within ~3 – 5 ps, making any possible spectral deformation negligible [7,8].

## S3. Extended data sets

In this Section we report extended data sets taken at selected pump-probe time delays, for various sample temperatures and applied magnetic fields (Fig. S3.1-6). For each set, we show three different quantities: the reflectivity, $R(\omega)$, the real conductivity, $\sigma_1(\omega)$, and the imaginary conductivity, $\sigma_2(\omega)$, of the photoexcited material. Experimental data (colored circles) are displayed along with fits performed either with a model describing the optical response of a Josephson plasma or with a Drude model for metals (colored solid lines, see also Section S4 for fitting procedure). The corresponding equilibrium spectra are also reported in each panel as gray lines.

Note that all these data and fit results have been used to produce Fig. 3 and Fig. 4 of the main text, in which we show the dynamical evolution of the interlayer phase correlation length and of the energy loss function, respectively. For this reason, we redirect the reader to the main manuscript for an extensive discussion on the experimental data.

An exception is given by the spectra of Fig. S3.5, which were taken on LBCO 11.5% at $T$ = 30 K and $H$ = 7 T. These data fully overlap in fact with those taken on the same material at $T$ = 30 K and zero magnetic field (Fig. S3.4), and for this reason they are not reported in the main text.

The absence of any appreciable magnetic field dependence in the transient response at $T$ = 30 K may relate to the fact that the equilibrium stripe order is strengthened by a magnetic field only in presence of a superconducting phase (*i.e.* for $T < T_C$ and not for $T$ = 30 K). Therefore, the data of Fig. S3.5 is an additional indication that optically-enhanced (induced) superconductivity in LBCO correlates with the strength of the equilibrium stripe order.

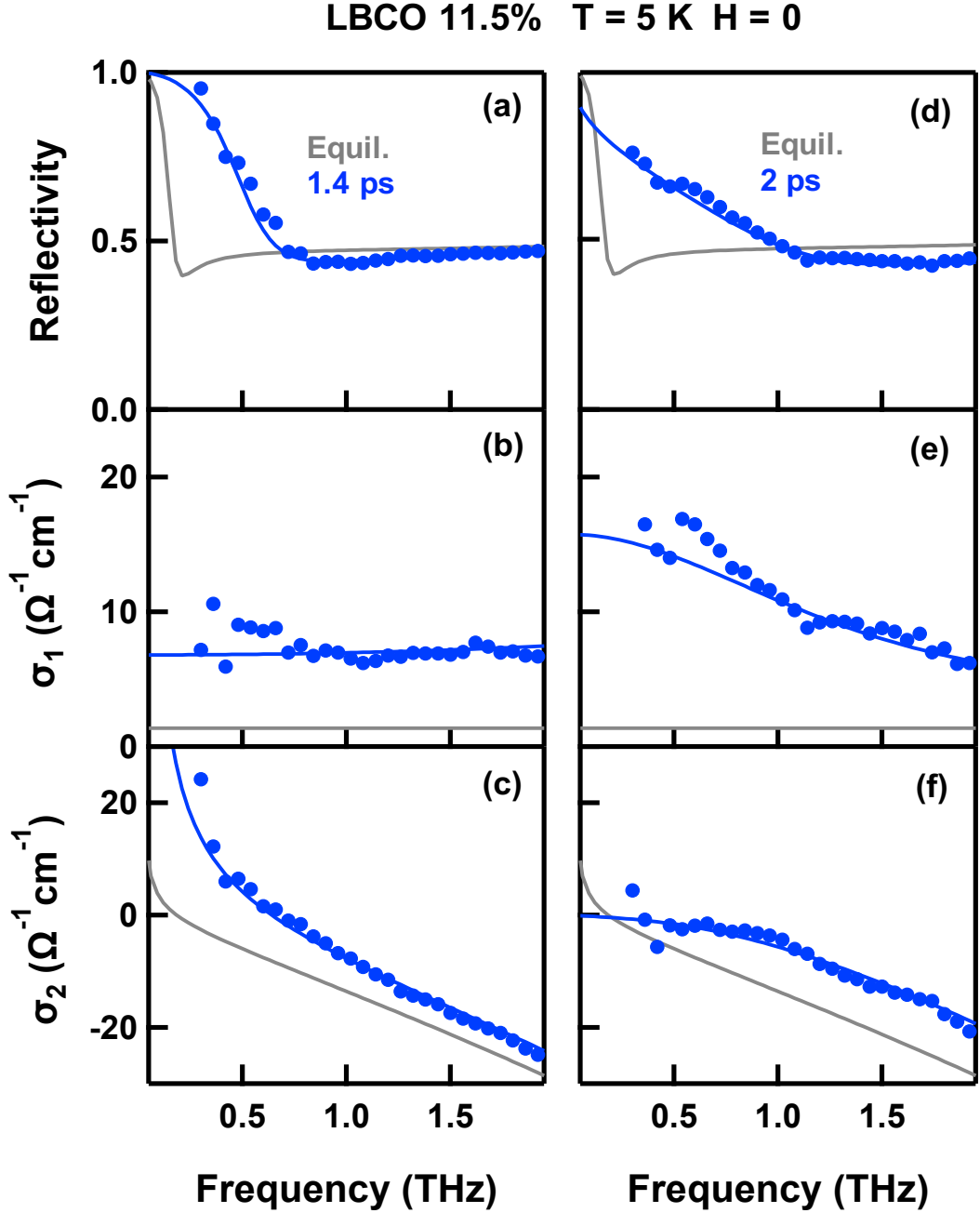

**Figure S3.1.** (a) *c*-axis THz reflectivity, (b) real, and (c) imaginary part of the optical conductivity of La$_{1.885}$Ba$_{0.115}$CuO$_4$ measured at *T* = 5 K (no magnetic field), at equilibrium (gray lines) and at $\tau$ = 1.4 ps pump-probe time delay (blue circles). Fits to the transient spectra are shown as blue lines. (d,e,f) Same quantities as in (a,b,c), measured at equilibrium (gray lines) and at $\tau$ = 2 ps time delay (blue). The transient data in (a,b,c) were fitted with a model describing the optical response of a Josephson plasma, while those in (d,e,f) with a Drude model for metals (see Section S4).

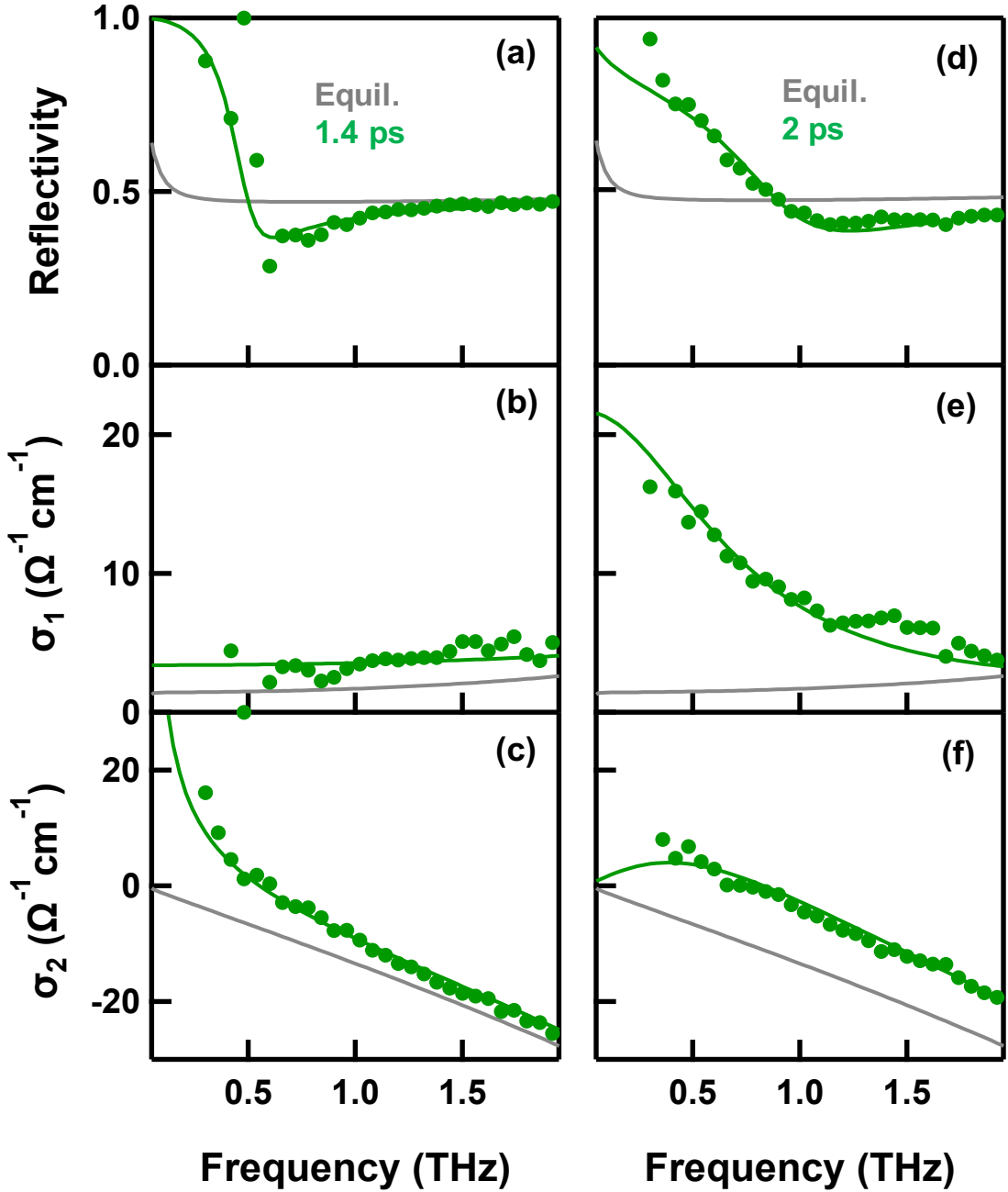

**Figure S3.2.** (a) *c*-axis THz reflectivity, (b) real, and (c) imaginary part of the optical conductivity of $La_{1.885}Ba_{0.115}CuO_4$ measured at $T = 5$ K in a 3.5 T magnetic field, at equilibrium (gray lines) and at $\tau = 1.4$ ps pump-probe time delay (green circles). Fits to the transient spectra are shown as green lines. (d,e,f) Same quantities as in (a,b,c), measured at equilibrium (gray lines) and at $\tau = 2$ ps time delay (green). The transient data in (a,b,c) were fitted with a model describing the optical response of a Josephson plasma, while those in (d,e,f) with a Drude model for metals (see Section S4).

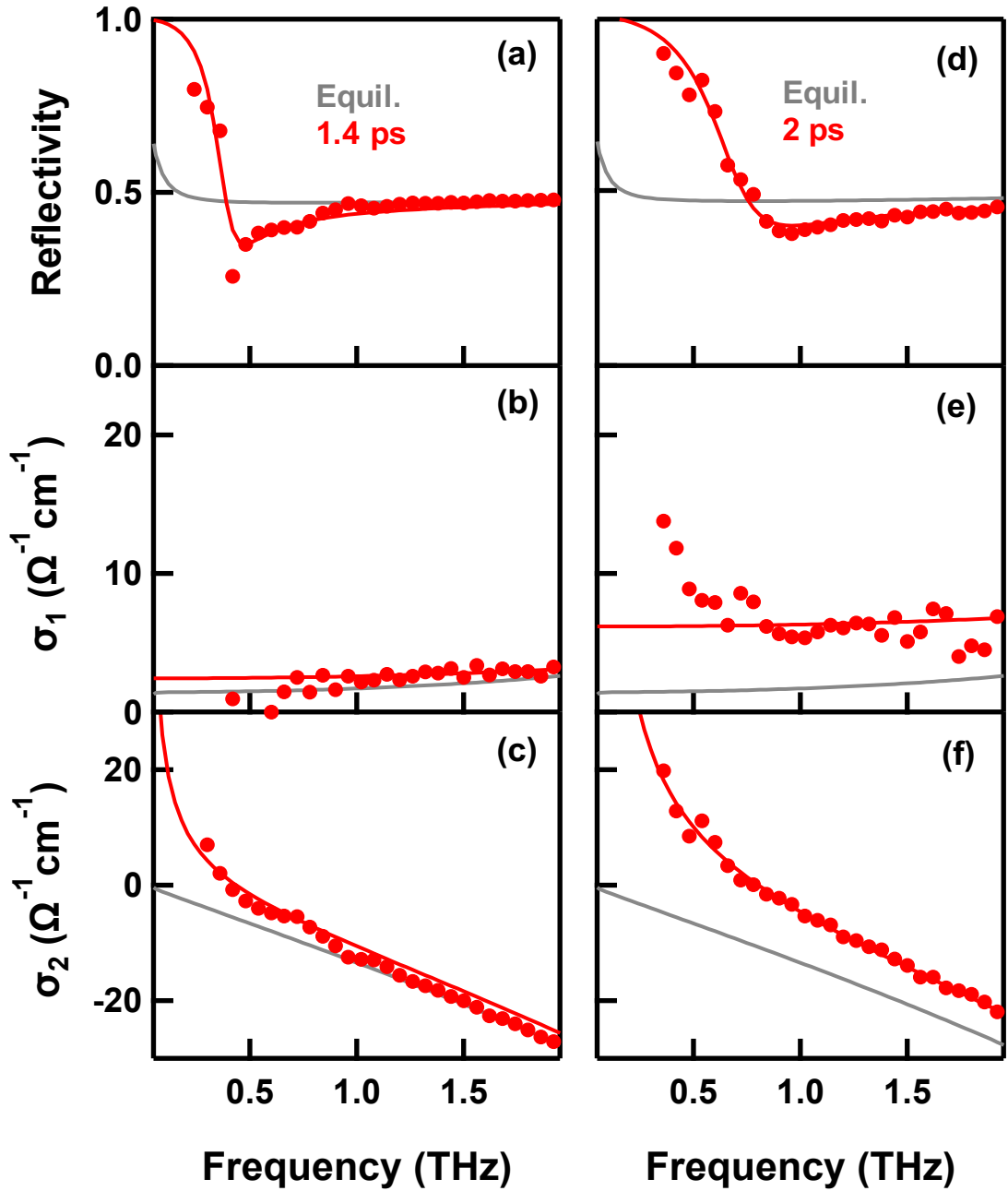

**Figure S3.3.** (a) *c*-axis THz reflectivity, (b) real, and (c) imaginary part of the optical conductivity of La$_{1.885}$Ba$_{0.115}$CuO$_4$ measured at *T* = 5 K in a 7 T magnetic field, at equilibrium (gray lines) and at *τ* = 1.4 ps pump-probe time delay (red circles). Fits to the transient spectra are shown as red lines. (d,e,f) Same quantities as in (a,b,c), measured at equilibrium (gray lines) and at *τ* = 2 ps time delay (red). All transient data were fitted with a model describing the optical response of a Josephson plasma (see Section S4).

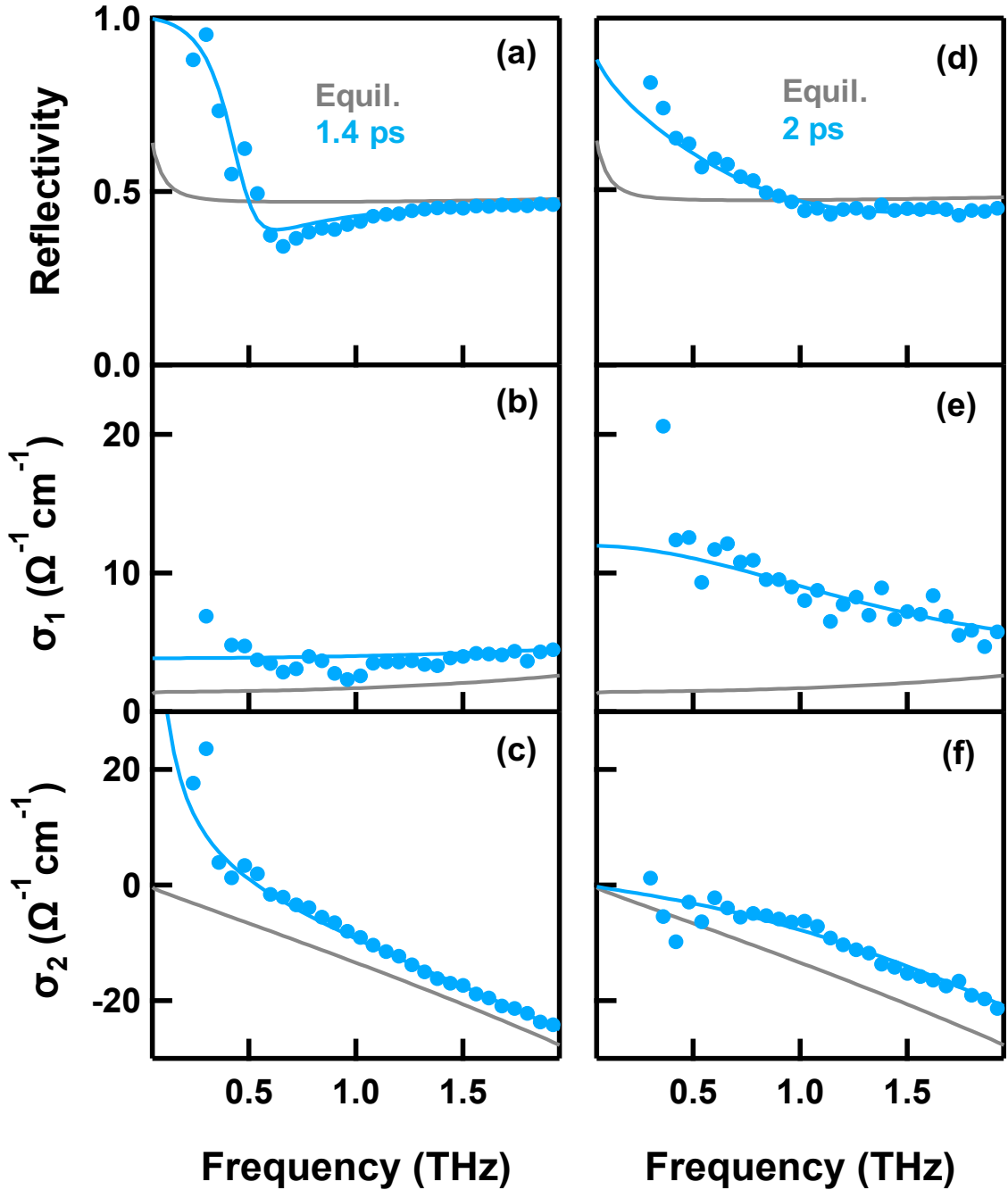

**Figure S3.4.** (a) *c*-axis THz reflectivity, (b) real, and (c) imaginary part of the optical conductivity of La$_{1.885}$Ba$_{0.115}$CuO$_4$ measured at $T$ = 30 K (no magnetic field), at equilibrium (gray lines) and at $\tau$ = 1.4 ps pump-probe time delay (light blue circles). Fits to the transient spectra are shown as light blue lines. (d,e,f) Same quantities as in (a,b,c), measured at equilibrium (gray lines) and at $\tau$ = 2 ps time delay (light blue). The transient data in (a,b,c) were fitted with a model describing the optical response of a Josephson plasma, while those in (d,e,f) with a Drude model for metals (see Section S4).

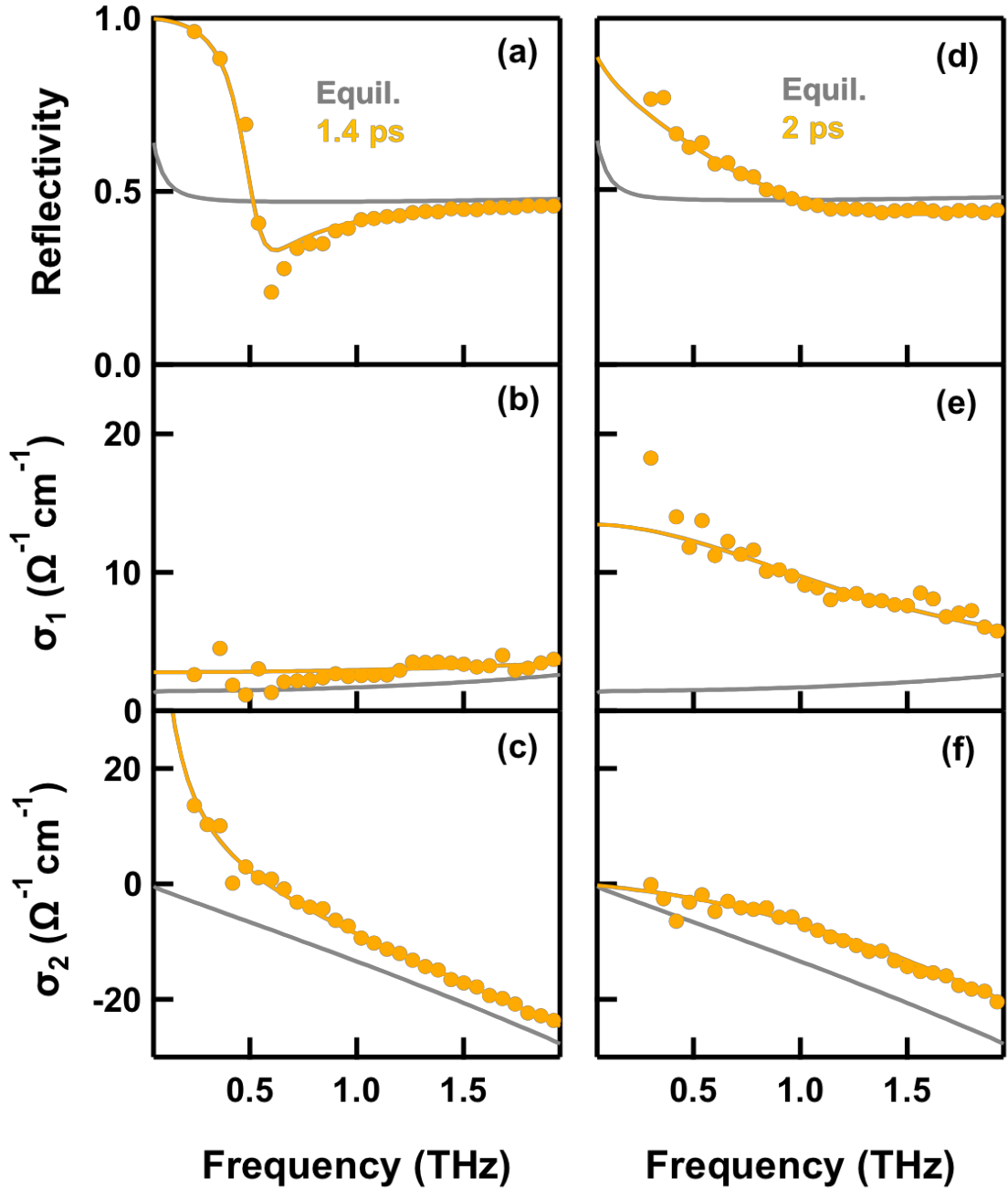

**Figure S3.5.** (a) *c*-axis THz reflectivity, (b) real, and (c) imaginary part of the optical conductivity of La$_{1.885}$Ba$_{0.115}$CuO$_4$ measured at *T* = 30 K in a 7 T magnetic field, at equilibrium (gray lines) and at $\tau$ = 1.4 ps pump-probe time delay (yellow circles). Fits to the transient spectra are shown as yellow lines. (d,e,f) Same quantities as in (a,b,c), measured at equilibrium (gray lines) and at $\tau$ = 2 ps time delay (light blue). The transient data in (a,b,c) were fitted with a model describing the optical response of a Josephson plasma, while those in (d,e,f) with a Drude model for metals (see Section S4).

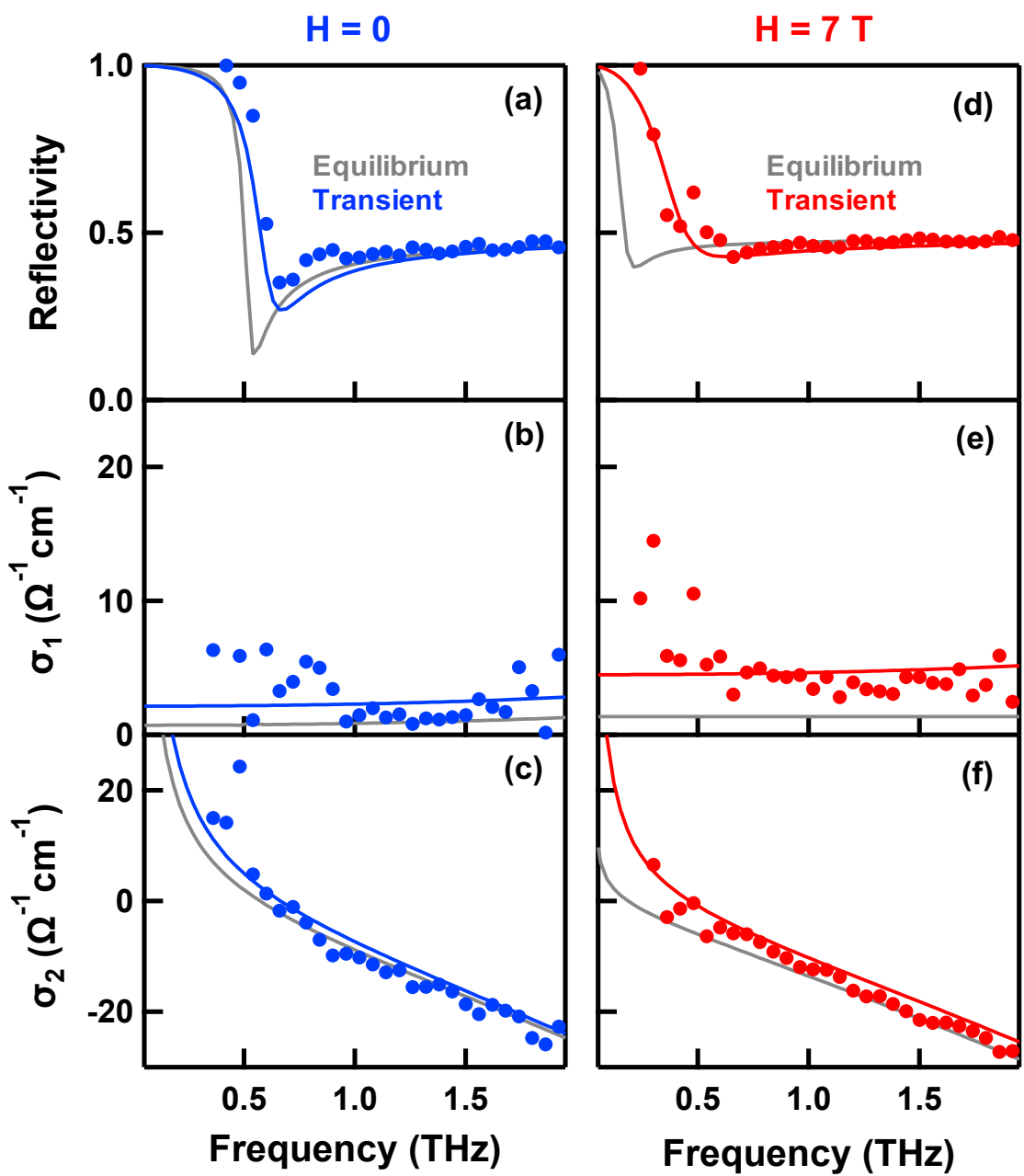

**Figure S3.6.** (a) *c*-axis THz reflectivity, (b) real, and (c) imaginary part of the optical conductivity of $La_{1.905}Ba_{0.095}CuO_4$ measured at *T* = 5 K (no magnetic field), at equilibrium (gray lines) and at $\tau$ = 1 ps pump-probe time delay (blue circles). Blue lines show fits to the transient spectra with a model describing the response of a Josephson plasma. (d,e,f) Same quantities as in (a,b,c), measured in $La_{1.905}Ba_{0.095}CuO_4$ at *T* = 5 K in presence of a 7 T magnetic field. Transient data, taken at $\tau$ = 1.4 ps, are shown here as red circles, while fits with the Josephson plasma model as red lines.

## S4. Fitting procedure

All transient optical spectra reported in this work could be satisfactorily reproduced using either a model describing the response of a Josephson Plasma or with a Drude model for metals. For each given data set (taken at a specific doping, sample temperature, magnetic field value and pump-probe time delay $\tau$), the transient reflectivity, $R(\omega)$, the real part of the optical conductivity, $\sigma_1(\omega)$, and its imaginary part, $\sigma_2(\omega)$, were simultaneously fitted with a single set of parameters.

The phonon modes in the far- and mid-infrared spectral region (5 THz ≲ $\omega$ ≲ 20 THz) and the high-frequency electronic absorption ($\omega$ ≳ 100 THz) were fitted from the equilibrium spectra (Fig. 1(c) of main text) with Lorentz oscillators, for which the complex dielectric function is expressed as

$$\tilde{\varepsilon}_{HF}(\omega) = \sum_i \frac{S_i^2}{(\Omega_i^2 - \omega^2) - i\omega\Gamma_i},$$

and kept fixed throughout the whole analysis. Here, $\Omega_i$, $S_i$, and $\Gamma_i$ are central frequency, strength, and damping coefficient of the $i^{th}$ oscillator, respectively.

The low-frequency spectral range ($\omega$ ≲ 2.5 THz), which was directly probed in our pump-probe experiment, required the introduction of an additional term in the model. For all data taken at long time delays ($\tau$ ≳ 2 ps), for which the reflectivity edge appeared broadened, each set of transient optical spectra could be well reproduced by a simple Drude term (see Fig. S4(g-i)). The full complex dielectric function used in this case is expressed as

$$\tilde{\varepsilon}_D(\omega) = \varepsilon_\infty[1 - \omega_P^2/(\omega^2 + i\Gamma\omega)] + \tilde{\varepsilon}_{HF}(\omega)$$

where $\omega_P$ and $\Gamma$ are the Drude plasma frequency and momentum relaxation rate, which were left as free parameters for the fit, while $\varepsilon_\infty$ was kept fixed to 4.5, a standard value for high-$T_C$ cuprates [3].

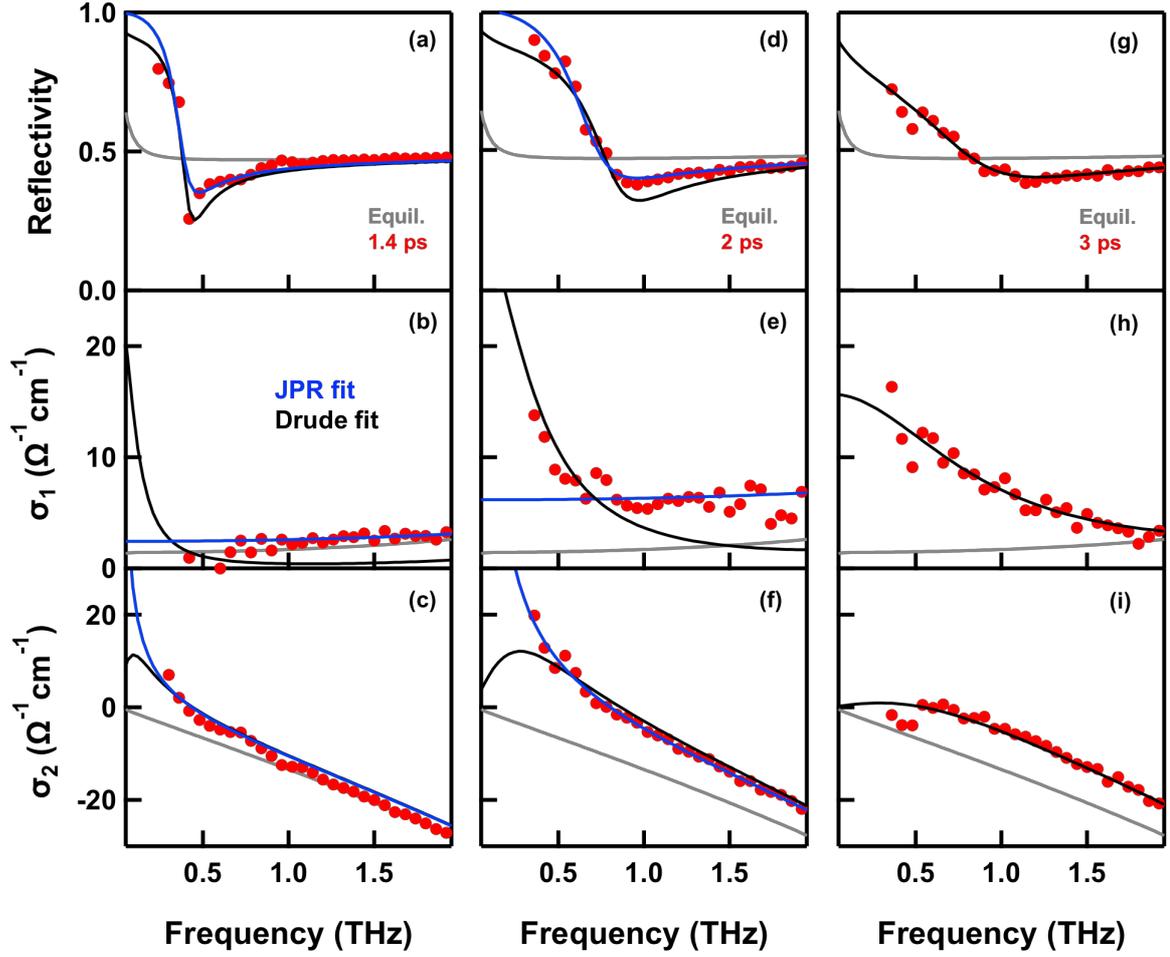

**Figure S4.** (a) *c*-axis THz reflectivity, (b) real, and (c) imaginary part of the optical conductivity of $La_{1.885}Ba_{0.115}CuO_4$ measured at $T$ = 5 K in a 7 T magnetic field, at equilibrium (gray lines) and at $\tau$ = 1.4 ps pump-probe time delay (red circles). Best fits to the transient spectra performed with a Josephson plasma and with a Drude model are shown as blue and black lines, respectively. (d,e,f) Same quantities as in (a,b,c), measured at equilibrium (gray lines) and at $\tau$ = 2 ps time delay (red circles). (g,h,i) Same quantities as in (a,b,c), measured at equilibrium (gray lines) and at $\tau$ = 3 ps time delay (red circles). Here, no Josephson plasma model was used to reproduce the data.

The transient response at earlier time delays ($\tau \lesssim 2$ ps) could also be satisfactorily reproduced by the same Drude model introduced above (black curves in Fig. S4(a-f)), although the momentum relaxation rate converged to values below the lowest probed frequency ($\Gamma < 0.3$ THz).

Alternatively, fits of at least comparable quality were performed, for these early delays, using a model describing the optical response of a Josephson Plasma at equilibrium (blue curves in Fig. S4(a-f)). Such model, which was also employed to reproduce the equilibrium spectra in the superconducting state (see Fig. S1), is able to fully capture all main features observed in the experimental data at early delays, *i.e.*, a sharp edge in $R(\omega)$, a gapped $\sigma_1(\omega)$, and a diverging $\sigma_2(\omega)$ toward low frequencies.

The full dielectric function is expressed in this case as

$$\tilde{\varepsilon}_J(\omega) = \varepsilon_\infty \big(1 - \omega_J^2/\omega^2\big) + \tilde{\varepsilon}_N(\omega) + \tilde{\varepsilon}_{HF}(\omega)$$

Here, the free fit parameters are the Josephson Plasma frequency, $\omega_J$, and $\tilde{\varepsilon}_N(\omega)$, a weak "normal fluid" component [9] (overdamped Drude term), which was introduced to reproduce the small positive offset in $\sigma_1(\omega)$ (see Fig. S4(b,e)).

Importantly, both models discussed here consistently return optical properties compatible with a superconducting response for frequencies above $\omega \gtrsim \Gamma$, with $\Gamma$ playing here the role of an inverse lifetime, $1/\tau_L$, of the transient superconducting state. This means that, within the frequency range over which the optical response of the transient state can be defined ($\omega \gtrsim \Gamma = 1/\tau_L$), a transient superconductor with lifetime $\tau_L$ and a normal conductor with momentum relaxation rate $\Gamma = 1/\tau_L$ are undistinguishable.

Despite this intrinsic ambiguity, we stress here that the evidence of a transient superconducting state in our measurement is very solid. Firstly, it should be emphasized that, starting from a weak superconductor at equilibrium (as indicated by a pre-existing Josephson Plasma Resonance, see *e.g.* Fig. 4(c) in main text), we observe a continuous transformation into a qualitatively identical phase with a blue-shifted resonance. At least for early time delays ($\tau \lesssim 2$ ps), the width of this resonance is essentially limited by our frequency resolution, returning $1/\Gamma$ values in excess of ~5-10 ps, from which one may extract carrier mobilities of at least $\sim 10^4$ cm$^2$/(V·s). These are extremely high values, which would be unprecedented for out-of-plane transport in a highly resistive normal state oxide [10].

Even at later time delays ($\tau \gtrsim 2$ - 3 ps), the momentum relaxation rates returned by Drude fits, $1/\Gamma \approx 1$ ps, are still anomalously high for conventional incoherent charge transport, and are rather suggestive of a strongly fluctuating superconducting state [11,12], which may persist for several picoseconds after photo-excitation.

Finally, the fact that such high mobility transport occurs at a density which is very close to the *c*-axis density of Cooper pairs in the same compound at equilibrium (as determined by the frequency of the plasma resonance) gives additional evidence in support of the picture of optically enhanced superconducting transport in photo-stimulated La$_{2-x}$Ba$_x$CuO$_4$.

# REFERENCES (Supplemental Material)